\documentclass[prl,aps,amssymb,showpacs,twocolumn]{revtex4}
\usepackage{amsmath}
\usepackage{amssymb}
\usepackage{amsthm}
\usepackage{amsfonts}
\usepackage{enumerate}
\usepackage{latexsym}
\usepackage[dvips]{graphicx}

\newcommand{\beq}{\begin{equation}}
\newcommand{\eneq}{\end{equation}}

\input{epsf}

\begin{document}

\tolerance 10000


\title{Intrinsic Spin-Hall Effect in n-Doped Bulk GaAs}

\author {B. Andrei Bernevig and Shou-Cheng Zhang}

\affiliation{Department of Physics, Stanford University, Stanford,
CA 94305}
\begin{abstract}
\begin{center}

\parbox{14cm}{We show that the bulk Dresselhauss ($k^3$)
spin-orbit coupling term leads to an intrinsic spin-Hall effect in
n-doped bulk GaAs, but without the appearance of uniform
magnetization. The spin-Hall effect in strained and unstrained bulk
GaAs has been recently observed experimentally by Kato {\emph{et.
al.}} \cite{kato2004A}. We show that the experimental result is
quantitatively consistent with the intrinsic spin-Hall effect due to
the Dresselhauss term, when lifetime broadening is taken into
account. On the other hand, extrinsic contribution to the spin-Hall
effect is several orders of magnitude smaller than the observed
effect.}
\end{center}
\end{abstract}

\maketitle

Recent theoretical work predicts dissipationless spin currents
induced by an electric field in semiconductors with spin-orbit
coupling\cite{murakami2003,sinova2003}. The response equation is
given by $j_j^i = \sigma_s \epsilon_{ijk} E_k$, where $j_j^i$ is the
current of the $i$-th component of the spin along the direction $j$
and $\epsilon_{ijk}$ is the totally antisymmetric tensor in three
dimensions. The response equation was derived by Murakami, Nagaosa
and Zhang\cite{murakami2003} for p-doped semiconductors described by
the Luttinger model of the spin-$3/2$ valence band. In another
proposal by Sinova \emph{et al.} \cite{sinova2003}, the spin current
is induced by an in-plane electric field in the 2-dimensional
electron gas (2DEG) described by the Rashba model\cite{sinova2003}.
The intrinsic spin-Hall effect predicted by these recent theoretical
works is fundamentally different from the extrinsic spin-Hall effect
\cite{d'yakonov1971,hirsch1999} due to the Mott type of skew
scattering by impurities \cite{mott1929}. The intrinsic spin-Hall
effect arises from the spin-orbit coupling of the host semiconductor
band, and has a finite value in the absence of impurities. On the
other hand, the extrinsic spin-Hall effect arises purely from the
spin-orbit coupling to the impurity atoms.

Experimental observation of the spin-Hall effect has been recently
reported by Kato \emph{et al.} \cite{kato2004A} in an electron doped
bulk sample and by Wunderlich \emph{et al.} in a two dimensional
hole gas (2DHG)\cite{wunderlich2004}. The 2DHG experiment has been
analyzed in a previous paper \cite{bernevig2004A} where it was shown
that the vertex correction due to potential impurity scattering
vanishes for that particular system. The experimental system is also
in the regime where lifetime broadening due to impurity scattering
is much less than the spin splitting, thus strongly suggesting an
intrinsic mechanism of the spin-Hall effect. In the experiment of
Ref. \cite{kato2004A}, spin accumulation due to a spin current is
observed even in the unstrained GaAs where no apparent spin
splitting is observed. The absence of observed spin splitting seems
to show the absence of intrinsic spin-orbit coupling in unstrained
n-doped GaAs. This fact prompted the authors of Ref.
\cite{kato2004A} to interpret the observed spin-Hall effect in terms
of the extrinsic mechanism due to impurity scattering only. In this
paper we show that, under close scrutiny, the results of
\cite{kato2004A} are consistent with an intrinsic mechanism. We
first show that the unstrained GaAs has a Dresselhauss $k^3$ spin
splitting which escapes detection by the method used in
\cite{kato2004A,kato2004B}. We then show that this spin splitting
leads to a spin-Hall current. This therefore explains the observed
spin accumulation on the edges of the unstrained GaAs within the
framework of the intrinsic spin-Hall effect. Furthermore, the
observed magnitude is consistent with the theory, after lifetime
broadening due to impurity scattering is taken into account. We also
predict that the bulk Dresselhauss term produces no net uniform
magnetization in the sample, this being generated solely by the
strain terms. In the case of strained GaAs, we compute the
self-energy correction in the weak spin-orbit coupling limit and
find a value for the spin-Hall conductivity close (enough) to the
measured value. The independence of the spin current on the
crystalographic directions can also be explained by the dominance of
the $k^3$ term over the $k$-linear terms induced by the small
strain. We also perform an order-of magnitude estimate and find out
that the extrinsic spin-Hall effect is seven orders of magnitude
lower than the clean limit of the intrinsic spin-Hall effect, and
several orders of magnitude lower than the observed experimental
value.

Let us first examine the extrinsic spin-Hall effect. In the
extrinsic mechanism\cite{d'yakonov1971,hirsch1999}, there is no
spin-orbit coupling in the band structure, and the spin-Hall effect
is caused by the scattering of electrons by the spin-orbit
interaction with impurities. The Hamiltonian is given by:
\begin{equation}
H = \frac{\hbar^2 k^2}{2m} + \frac{ \hbar^2}{2m^2 c^2} \vec{\sigma}
(\vec{\nabla} V(r) \times \vec{k}) \label{extrinsicH},
\end{equation}
\noindent where $V(r)$ is the impurity potential. The extrinsic
spin-Hall effect is basically derived from the atomic Mott
scattering \cite{mott1929}, and the important length scale is
governed by the Compton wave length $\lambda_c=\hbar/mc$. The
extrinsic spin-Hall effect has been computed systematically for this
Hamiltonian\cite{crepieux2001}, and the order of the magnitude of
the effect can be estimated to be:
\begin{equation}
\sigma_{extrinsic} \sim \frac{e^2}{\hbar} (\lambda_c k_F)^2 k_F
\label{extrinsicV},
\end{equation}
where $k_F$ is the fermi wave vector. For the experimental system by
Kato \emph{et al.}, with $k_F =10^8 m^{-1}$ and a conduction band
effective mass of $m=0.0665 m_e$, Eq. (\ref{extrinsicV}) gives
$\sigma_{extrinsic} = 1.2 \times 10^{-4} \Omega^{-1} m^{-1}$, almost
$4$ orders of magnitude smaller than the observed spin-Hall
conductance. On the other hand, the intrinsic spin-Hall effect is a
genuine solid state effect, governed purely by the fermi wave vector
$k_F$, and the order of magnitude of the effect is given by
\begin{equation}
\sigma_{intrinsic}^{clean} \sim \frac{e^2}{\hbar} k_F
\label{intrinsicV}
\end{equation}
in the clean limit. Therefore, we see that the ratio of the two
effects is given by \cite{murakami2003A}
\begin{equation}
\frac{\sigma_{extrinsic}}{\sigma_{intrinsic}^{clean}} \sim
(\lambda_c k_F)^2 \sim 10^{-7}.
\end{equation}
Therefore, in distinguishing between the two effects, it is
extremely important to keep in mind the smallness of the
dimensionless parameter $\lambda_c k_F$. In the literature of the
anomalous Hall effect, a so called ``enhancement factor" is
sometimes introduced in a rather ad-hoc
basis\cite{berger1970,crepieux2001}. However, this ``enhancement
factor" is microscopically based on the spin-orbit coupling within
the band structure, and would necessarily lead to a spin splitting
of the bands. Therefore, we can safely conclude that {\it if there
were no spin splitting due to the intrinsic spin-orbit coupling
within the band}, the extrinsic spin-Hall effect is far too small
to explain the experiment by Kato \emph{et al.} in the unstrained
GaAs.

Let us now turn to the intrinsic spin-orbit coupling within the
conduction band. The Hamiltonian of an inversion asymmetric bulk
(unstrained) semiconductor contains a Dresselhauss $k^3$ spin
splitting term in the conduction band, which can be written as a
momentum dependent magnetic field:
\begin{equation} \label{unstrainedGaAsHamiltonian}
H=\frac{\hbar^2}{2m} k^2 + B_i({\bf{k}}) \sigma^i, \;\;\; i=1,2,3
\end{equation}
\noindent where $B_x= \gamma k_x (k_z^2 - k_y^2)$, $B_y = \gamma k_y
(k_x^2 - k_z^2)$, ${B_z = \gamma k_z (k_y^2 - k_x^2)}$. The coupling
constant $\gamma$ has been determined in a number of independent
experiments, and a value of $\gamma \approx 25 eV {\AA}^3$ is widely
quoted in the literature
\cite{dresselhauss1992,hassenkam1997,jusserand1995,knap1996}. We
must now reconcile this spin splitting with the fact that the
measurement carried out in Ref. \cite{kato2004B} does not see any
splitting in the unstrained sample. In \cite{kato2004B}, a spin
packet injected at the Fermi momentum is subsequently dragged by an
external electric field $\vec{E}$. Experiments are performed along
two crystallographic directions $\vec{E} || [110]$ and $\vec{E} ||
[1\bar{1}0]$. This creates an average nonzero particle momentum
$\langle \vec{k}\rangle \sim \frac{e}{\hbar} \vec{E} \tau$ which in
turn creates a non-zero average (over the Fermi surface) internal
magnetic field $\langle \vec{B} \rangle $. The spin splitting in
\cite{kato2004B} is obtained as a derivative of the averaged
$\langle \vec{B} \rangle $ with respect to the drag momentum
$\langle \vec{k} \rangle$. Due to the special symmetry of the
Dresselhauss spin-orbit coupling, this procedure turns out to yield
a null result, even if $\gamma$ is finite. Take, for example
$\vec{E} || [110]$, then the momentum of a particle injected near
the Fermi momentum $\vec{k}^F$ is:
\begin{equation}
\vec{k} = \vec{k}^F + \langle \vec{k} \rangle, \;\;\; \langle
\vec{k} \rangle = - \frac{e \tau}{m} \vec{E} ||[110], \;\;\; \langle
k_x \rangle =  \langle k_y \rangle.
\end{equation}
\noindent To first order in $\langle \vec{k} \rangle$, the
components (say $x$) of $\langle \vec{B} \rangle $ averaged over the
Fermi surface is:
\begin{equation}
 \langle B_x \rangle = \langle k_x \rangle \int
 \frac{d\Omega}{4 \pi}
\left[ (k_z^F)^2 - (k_y^F)^2 - 2 k^F_x k^F_y \right].
\end{equation}
\noindent Since the spin-orbit coupling term is much smaller than
the kinetic term, the Fermi surface is, to first order in $\gamma$,
a sphere (there is, of course, a zero order in $\langle k_x \rangle$
term, but this obviously vanishes upon integration over the Fermi
surface so we have omitted it). As the integration is carried over a
sphere, it is obvious that $\int (k_z^F)^2 = \int (k_y^F)^2 =
(k^F)^2/3$ and $\int k^F_x k^F_y =0$. Therefore $\langle B_x \rangle
=0$, and no spin splitting is expected from this procedure, even
though in the Dresselhauss term $\gamma$ maybe finite. Note that
this cancellation would not happen if the spin-orbit coupling term
were $k$-linear since the derivative of $\vec{B}_{int}$ would just
be a constant. This is exactly what happens in the strained samples
of GaAs where the spin splitting was explained by $k$-linear terms
\cite{bernevig2004}.

A related fact shows that, due to its symmetry, the bulk-Dresselhaus
term produces no uniform magnetization in the bulk of the sample.
This is an easily falsifiable prediction of our theory. The
Hamiltonian (\ref{unstrainedGaAsHamiltonian}) has two energy levels
$E_\pm = \frac{\hbar^2}{2m} k^2 \pm B$ where $B=\sqrt{B_i B_i}$. The
uniform magnetization $\langle \sigma_i \rangle$ induced by an
electric current $J_j = \partial H / \partial k_j$ (due to the
applied electric field $E_j$) can be easily computed in linear
response, and one obtains:
\begin{eqnarray}
& \langle \sigma_i \rangle= \frac{2 \pi e \tau}{\hbar} Q_{ij}  E_j
\nonumber  \\ & Q_{ij} = \langle T \sigma_i J_j \rangle = \int
\frac{d^3 k}{(2 \pi)^3} \frac{n_{E_-} - n_{E_+}}{B^2} \left( B_i
\frac{\partial B}{\partial k_j} - B \frac{\partial B_i}{\partial
k_j} \right)
\end{eqnarray}
\noindent where $n_{E_\pm}$ are the Fermi functions of the two
bands. By inspection, all the components of $B_i \frac{\partial
B}{\partial k_j} - B \frac{\partial B_i}{\partial k_j}$ are odd in
the components $k_i$ and hence vanish under integration due to cubic
symmetry. This leads to $\langle \sigma^i \rangle \equiv 0$. By
contrast, a $k$-linear internal magnetic field, as in the strained
samples, gives a finite uniform magnetization due to the fact that
$\frac{\partial B_i}{\partial k_j}$ is a constant while $B$ is
isotropic in (and proportional to) $k$ \cite{bernevig2004}. The bulk
Dresselhauss term is most likely also the explanation of the
contradiction between the observed spin splitting along the $[110]$
and $[1\bar{1} 0]$ directions, and the uniform magnetization on
these directions. In \cite{kato2004} it is observed that although
the spin splitting for $E ||[110]$ is consistently larger than the
spin splitting for the $E || [1\bar{1} 0]$, the uniform
magnetization for $E || [110]$ is usually lower than that for $E ||
[1\bar{1} 0]$. This can be very well explained by the missing
bulk-Dresselhauss term in the case when this term subtracts from the
splitting on the $[110]$ direction but adds to the splitting in the
$[1\bar{1}0]$ directions, which makes it qualitatively possible that
the uniform magnetization observed be in agreement with the spin
splitting. Quantitative modelling of this involves precise knowledge
of the sample Hamiltonian (including the strain) which is currently
not possible (mainly because of limited knowledge about out-of-plane
spin-orbit coupling terms).

Even though it creates no uniform magnetization, the bulk
Dresselhauss term does give rise to an intrinsic spin-Hall effect
when under the action of an electric field. We define the spin
current as usual ($\varepsilon = \hbar^2 k^2 /2m$):
\begin{equation}
J^l_i = \frac{1}{2} \left\{ \frac{\partial H}{\partial k_i},
\sigma^l \right\} = \frac{\partial \varepsilon}{\partial k_i}
\sigma_l + \frac{\partial B_l}{\partial k_i},
\end{equation} \noindent and the expanded expression for the Green's function $G(k, i\omega_n) = [i
\omega_n -H]^{-1}$ as:
\begin{eqnarray}
& G(k, i\omega_n)= f(k, i\omega_n) (g(k, i \omega_n) + B_i(k)
\sigma_i) \nonumber \\ &
 f(k, i\omega_n) = \frac{1}{(i\omega_n  -
\varepsilon(k))^2 - B^2},\;\; g(k, i\omega_n) = i\omega_n -
\varepsilon(k).
\end{eqnarray}
\noindent When subjected to the action of an electric field
$\vec{E}$, the frequency dependent spin conductance (not including
the vertex correction) can be found in linear response as:
\begin{eqnarray}
& J_i^l = \sigma_{ij}^l E_j;\; \;\;\; \sigma^l_{ij} = \frac{
Q^l_{ij} (\omega) }{-i \omega} ; \;\;\;\; Q^{l}_{ij} (i \nu_m)=
\nonumber \\ & = \frac{1}{V \beta} \sum_{k, n} Tr[G(k, i(\omega_n +
\nu_m)) J^l_i (k) G(k, i\omega_n) J_j(k)].
\end{eqnarray} \noindent
Summing over the Matsubara frequencies  $i \omega_n$, analytically
continuing $i \nu_m \rightarrow \omega$, as well as omitting a
dissipative term which vanishes upon momentum integration, we
obtain the expression for the frequency dependent (reactive) spin
conductivity:
\begin{equation} \label{dresselhaussspinconductivity}
\sigma_{ij}^l (\omega) = \frac{\hbar^2 }{2m}\int \frac{d^3 k}{(2
\pi)^3} \frac{n_{E_-} - n_{E_+}}{B(B^2 - \omega^2) } k_i
\epsilon_{lnr} B_n \frac{\partial B_r}{\partial k_j}
\end{equation}
\noindent where $i,j,l,r,n =x,y,z$. Unlike the uniform magnetization
case, the integrand is even in $k$ and finite upon integration.
Hence the spin current in the unstrained GaAs can be qualitatively
explained by the presence of a intrinsic spin-Hall effect due to the
bulk-Dresselhauss term. Working in spherical coordinates,
substituting the explicit expression for the bulk Dresselhauss spin
splitting $B_i(k)$, using the expression (valid for small $\gamma$)
of the difference between the Fermi momenta of the two spin-split
bands:
\begin{equation}
k^F_- - k^F_+ \approx \frac{2 m}{\hbar^2} \frac{B(k^F)}{k^F}; \;\;\;
k^F = \frac{k^F_+ + k^F_-}{2},
\end{equation}
\noindent  as well as integrating over the spherical angles, one
obtains for the $DC$ spin-Hall conductivity:
\begin{equation}
\sigma^l_{ij} = \frac{k^F}{12 \pi^2} \epsilon_{lij}.
\end{equation}
\noindent This is the intrinsic spin-Hall conductivity in the clean
limit, $\hbar/\tau << B(k)$, which we call
$\sigma^{clean}_{intrinsic}$. For the carrier concentration in
\cite{kato2004A} we have $k^F = 10^8 m^{-1}$ and we obtain
$\sigma^{clean}_{intrinsic}= 200 \Omega^{-1} m^{-1}$. This is much
larger than the observed conductivity of $0.2 \sim 0.5 \Omega^{-1}
m^{-1}$. But this is expected since we have so far not taken the
influence of disorder into account . In the experiment by Kato
\emph{et al.}, the lifetime broadening due to impurity scattering is
much larger than the weak spin splitting due to the Dresselhauss
coupling. The intrinsic spin-Hall effect is therefore in the dirty
limit, and a significant reduction from the clean result is
therefore expected. Note that the lifetime broadening due to
impurity scattering leads to a {\it reduction} of the intrinsic
spin-Hall conductivity. This is different from the extrinsic
spin-Hall effect due to spin-orbit coupling to the impurity
potential, which makes a small, but {\it positive} contribution to
the spin-Hall conductivity.

To properly take into account disorder, one must perform a
self-consistent calculation taking into account both the self-energy
and the vertex correction. In the case of the electron Rashba model
with a $k$-linear spin splitting, many groups have shown that the
vertex correction cancels the intrinsic spin-Hall effect
\cite{inoue2004,mishchenko2004}. However, this cancellation seems to
be special to the $k$-linear spin splitting, and it has been shown
that the vertex correction due to $k^2$ light/heavy hole splitting
in the Luttinger model, or due to $k^3$ spin splitting in the heavy
hole band, vanishes identically \cite{murakami2004,bernevig2004A}.
We expect that for a similar reason, the vertex correction due to
the $k^3$ Dresselhauss spin splitting would not cancel the intrinsic
spin-Hall effect either. We will hence neglect the vertex correction
and focus on the self-energy correction which is easy to extract
analytically. The self-energy approximation to disorder can be
simulated by letting $\omega = i \hbar/\tau$ in Eq.
(\ref{dresselhaussspinconductivity}). The values in \cite{kato2004A}
are in the regime $\hbar/\tau >> B(k)$, we thus obtain:
\begin{equation}
\sigma_{ij}^l  = \frac{\hbar^2 }{2m} \frac{1}{(\hbar /\tau)^2}\int
\frac{d^3 k}{(2 \pi)^3} \frac{n_{E_-} - n_{E_+}}{B} k_i
\epsilon_{lnr} B_n \frac{\partial B_r}{\partial k_j}
\end{equation}
\noindent which, upon momentum integration gives the lower bound for
the spin conductivity:
\begin{equation}
\sigma_{ij}^k  =\frac{4 k_F}{105 \pi^2}\left(\frac{\gamma
k_F^3}{\hbar /\tau}\right)^2 \epsilon_{ijk}.
\end{equation}
\noindent For the values $\gamma = 25 eV {\AA}^3$, $k_F = 10^8
m^{-1}$ we obtain a bulk Dresselhauss spin splitting energy $\gamma
k_F^2 \approx 0.025 meV$ while $\hbar/\tau \approx 1.6 meV$ for a
sample of mobility $\mu =1 m^2/Vs$ as the one in the experiment.
Using these values, we obtain for the intrinsic, disorder quenched
spin conductivity $\sigma^{dirty}_{intrinsic} = 0.02 \Omega^{-1}
m^{-1}$. This lower bound is smaller than the measured conductivity
(which is $0.2 \Omega^{-1} m^{-1}$ for small electric field and $0.5
\Omega^{-1} m^{-1}$ for large electric field. This is a lower bound
for the spin conductivity since $\hbar/\tau$ is an upper bound for
the frequency $\omega$ in the dirty limit. Considering the
uncertainty associated with the value of $\gamma$, the crudeness of
the estimate, and the indirect determination of the experimental
value, the agreement is reasonably good.

The application of strain induces two extra spin splittings in the
Hamiltonian which are linear in the momentum $k$
\cite{bernevig2004}. There is one structural inversion asymmetry
(SIA) splitting of the form $\alpha (k_y \sigma_x - k_x \sigma_y)$,
and a bulk-inversion asymmetry (BIA) of the form $\delta(k_x
\sigma_x - k_y \sigma_y)$, where $\alpha$ and $\delta$ are strain
dependent. For the values of the splitting in sample $E$ used in
\cite{kato2004B} we have $\alpha/\hbar = 183 m/s$ and $\delta /
\hbar = 112 m/s$. We observe that the splitting at the Fermi
momentum is $0.011 meV$ for the SIA term and $0.007 meV$ for the BIA
term. By contrast, the $k^3$ Dresselhauss coupling is $0.025 meV$,
so it is likely that it will dominate (although not overwhelmingly)
the spin current. Moreover, a vertex correction computation for the
SIA or BIA term separately reveals that the spin current caused by
these terms vanishes upon the introduction of impurities
\cite{inoue2004} (exact numerical diagonalization results
\cite{nomura2004,nikolic2004} are, however, at odds with
\cite{inoue2004}), whereas a vertex calculation for a $k^3$ term
shows finite spin current \cite{bernevig2004A,murakami2004}. It is
therefore plausible that the bulk Dresselhauss term dominates the
spin-Hall transport even in the strained samples. This naturally
explains the independence of spin current on the crystallographic
directions of the applied electric field, since the bulk spin
conductivity for the Dresselhauss term is direction independent. The
experimental features observed can therefore be qualitatively
explained by an intrinsic mechanism.

In conclusion, we have shown that without any spin splitting in
the electron band, the extrinsic spin-Hall effect is far too small
to explain the experimentally observed value of spin-Hall
conductivity in \cite{kato2004A}. In order to definitively
determine the origin of the spin-Hall effect, we propose to carry
out similar experiments in materials without any known intrinsic
spin-orbit coupling, and a null result would give the definitive
proof that the extrinsic spin-Hall effect is far below the current
experimental sensitivity, and can not be the origin of the
spin-Hall effect observed in Ref. \cite{kato2004A}. We have shown
that the experimental results are consistent with the
interpretation of an intrinsic spin-Hall effect in terms of a bulk
Dresselhauss term in the unstrained sample. Furthermore, this
intrinsic spin-orbit coupling is consistent with the apparent
absence of spin splitting observed (in the unstrained samples) in
the spin drag experiment\cite{kato2004B}. We also predict that the
uniform magnetization in the unstrained bulk GaAs samples will be
close to zero due to the symmetry of the $k^3$ Dresselhaus term.
Since the experiment is carried out in a regime where the lifetime
broadening due to impurity scattering is large compared to the
spin splitting, the observed spin-Hall conductivity is
significantly reduced from the value in the clean limit. It is
argued that the $k$-linear terms in the strained GaAs samples
\cite{kato2004B, bernevig2004}, although crucial for the
appearance of a uniform magnetization, have a limited effect on
the spin current due to the dominance of the Dresselhauss term,
thereby qualitatively explaining the direction independence of the
spin-Hall effect observed in the experiment.

We would like thank Y.K. Kato and and D. Awschalom  for many useful
discussions. B.A.B. acknowledges support from the Stanford Graduate
Fellowship Program. This work is supported by the NSF under grant
numbers DMR-0342832 and the US Department of Energy, Office of Basic
Energy Sciences under contract DE-AC03-76SF00515.


\end{document}